\documentclass[twocolumn,showpacs,preprintnumbers,amsmath,amssymb,epsf]{revtex4}
\usepackage{graphicx}
\usepackage{dcolumn}
\usepackage{bm}

\begin{document}

preprint{APS/123-QED}

\title{Updating schemes in zero-temperature single-spin flip dynamics}

\author{Sylwia Krupa and Katarzyna Sznajd--Weron}
\email{kweron@ift.uni.wroc.pl}
\homepage{http://www.ift.uni.wroc.pl/~kweron}
\affiliation{Institute of Theoretical Physics, University of
Wroc{\l}aw, pl. Maxa Borna 9, 50-204 Wroc{\l}aw, Poland }

\date{\today}

\begin{abstract}
In this paper we examine the role of the so called $c$-parallel updating schemes in relaxation from disordered states to the final ferromagnetic steady state. We investigate two zero-temperature single-spin flip dynamics on a one dimensional lattice of length $L$: inflow (i.e. generalized zero-temperature Glauber dynamics) and outflow opinion dynamics. The varying $c$ allows us to change the updating scheme from random sequential updating ($c=1/L$) to deterministic synchronous updating (for $c=1$). We show how the mean relaxation times depend on $c$ and scale with the system size $L$. Moreover, we empirically find an analytical formula for the ratio between mean relaxation times for inflow and outflow dynamics. Results obtained in this paper suggest that in some sense the original zero-temperature Glauber dynamics is a critical one among a broader class of inflow dynamics.    
\end{abstract}

\maketitle

\section{Introduction}

Recently, there has been renewed interest in the possibly different physics arising from the sequential or synchronous
execution of the microscopic update rule of the spins in disordered systems \cite{bolle_blanco_04} (and references therein).
Two most used updating schemes are - random sequential and synchronous (parallel) updating. Within synchronous updating all
units of the system are updated at the same time. Random sequential updating means that only one, randomly chosen, unit is updated
at each time step. Probably neither a completely synchronous nor a random sequential update is realistic for natural systems \cite{radicchi_etal_07}. Therefore, in the previous paper \cite{szn_kru_06} we introduced $c$-parallel updating, in which a randomly chosen fraction $c$ of spins is updated synchronously. Varying $c$ allows us to move from random sequential updating ($c=1/L$) to deterministic synchronous updating (for $c=1$). An analogous idea was proposed very recently by Radicchi et al. \cite{radicchi_etal_07} to study quench from infinite temperature to zero temperature of Ising spin systems evolving according to the Metropolis algorithm, in which the flipping probability associated with the $i$th spin is defined as \cite{mar_etal_94}:
\begin{equation}
p_M=\min \left\{ 1,\exp(-\Delta E_i/k_B T) \right\},
\end{equation} 
where $\Delta E_i$ is the energy change due to the $i$th spin flip. For $\Delta E_i=0$ we obtain $p_M=1$, hence zero-temperature limit of this algorithm can be rewritten as a special case of the so called generalized zero-temperature Glauber dynamics in which flips for $\Delta E_i>0$ are forbidden and the probability of flipping $i$th spin in case of $\Delta E_i=0$ is $W_0=1$ \cite{god_luc_05}. Another example of such a generalized dynamics (which we call in this paper inflow dynamics) is the zero-temperature limit of the original Glauber dynamics \cite{Glauber_63} -- the corresponding flipping probability is $W_0=1/2$. 

It occurred that for $W_0=1$ the critical value of $c$ exists above which the system never reaches the final ferromagnetic state \cite{radicchi_etal_07}. This critical value was determined by measuring the evolution of active bonds (bond is active if it connects two sites with opposite spins). We expect that in the case of generalized Glauber dynamics ($W_0 \in [0,1]$) the critical value of $c$ depends on $W_0$. However, determining precise dependence between the critical updating scheme $c$ and flipping probability $W_0$ is not the subject of this paper and will be considered in a future study. In the present paper we focus on measuring the mean relaxation times to the ferromagnetic steady state as a function of $c$ and $W_0$. Moreover, apart from the inflow dynamics, we investigate the so called outflow dynamics, which was originally proposed to study opinion dynamics \cite{szn_szn_00}. 

The paper is organized as follows. In the next section we present the motivation for this study. In the following sections we recall definitions of studied dynamics, as well as, the idea of $c$-parallel updating and present simulation results. We conclude the paper by summarizing the results.

\section{Motivation}

As Kenrick et al. \cite{KNC07} report, about 20 years ago, for no good reason the customers of a local bank in Singapore began drawing out their money in a frenzy. The run on this respected bank remained a mystery until much later, when researchers interviewing participants discovered its peculiar cause: an unexpected bus
strike had created an abnormally large crowd waiting at the bus stop in front of the bank that day. Mistaking the gathering for a crush of customers poised to withdraw their funds from a failing bank, passersby panicked and got in line to withdraw their deposits, which led more passersby to do the same. Soon illusion had become reality and, shortly after opening its doors, the bank was forced to close to avoid ruin.
This is only one of many examples of social validation described in social psychology textbooks \cite{KNC07,AWA07,Myers06}. 

Most people feel that behaviors become more valid when many others are performing them. Although the tendency to follow the lead of our peers can drive to misguided behavior, most of the time it sends us in right directions, toward correct choices. On the other hand, it happens that people often follow others even if they believe that the group may be in error. 
 
The research of Solomon Asch in 1955 demonstrated that, when faced with a strong group consensus, people often conform even
if they believe that the group may be in error \cite{Asch55}. However, even a single visible dissenter from the group's position emboldens others to resist conformity. A number of later experiments showed that an individual who breaks the unanimity principle reduces social pressure of the group dramatically \cite{Myers06}. This observation was recently expressed in a simple one dimensional USDF (`United we Stand, Divided we Fall') model of opinion formation \cite{szn_szn_00}. The model was later renamed `the Sznajd model' by Stauffer \cite{sta_etal_00} and generalized on a two dimensional square lattice.

The crucial difference between the Sznajd model and other Ising-type models of opinion dynamics \cite{galam_86,galam_90,kac_hol_96,hol_kac_sch_00,hol_kac_sch_01,kra_red_03} or zero-temperature Glauber dynamics \cite{Glauber_63} is that information flows outward from the center nodes to the surrounding neighborhood (so called outflow dynamics \cite{szn_kru_06}) and not the other way around. In spite of this difference, it has been suggested \cite{gal_05,beh_sch_03}, that outflow dynamics is equivalent with zero-temperature Glauber dynamics (inflow dynamics) for the Ising ferromagnet with the nearest neighbors interactions. 

The updating of an extended group of spins in the Ising model with zero-temperature Glauber kinetics \cite{Glauber_63} was considered by Newman and Stein \cite{new_ste_99}, Lipowski \cite{lipowski99} and later on by \cite{spi_kra_red_01a,spi_kra_red_01b,Jain02,kra_red_tai_04,bolle_blanco_04,god_luc_05,radicchi_etal_07,cha_das_08} (and referenced therein). 
Outflow dynamics in two dimensions was studied by Stauffer et al. \cite{sta_etal_00,mor_etal_00,ber_etal_01,sta_oli_02} and later on by others (for reviews see \cite{sta_02,sch_02,for_sta_05,sw_05,cas_for_lor_07}. Qualitative differences between outflow and inflow dynamics in two dimensions have been shown. Under outflow dynamics the system always relaxes to the ferromagnetic steady state. On the contrary, an Ising ferromagnet has a large number of metastable states with respect to the Glauber spin-flip dynamics. Therefore, at zero temperature the system could get stuck forever in one of these states \cite{spi_kra_red_01a}. On the other hand, since there are no metastable states in one dimension, the only possible final states are all spins up or all spins down, analogously to outflow dynamics.

Let us recall two differences between settings of inflow and outflow dynamics:
\begin{enumerate}
\item
Under zero-temperature Glauber (inflow) dynamics conformity is caused by majority, in contrast to outflow dynamics where unanimity is required. Of course, in one dimension, when a pair of spins is influencing others, majority is equivalent to unanimity -- this difference is important only in higher dimensions.
\item
Under inflow dynamics the center spin is influenced by its nearest neighbors, while under outflow dynamics information flows from the center spins to the neighborhood.
\end{enumerate}
Nevertheless, under random sequential updating, both dynamics seem to be equivalent in one dimension, except for one value of parameter $W_0 = 0$, which describes the probability of flipping the spin in lack of majority. This particular value $W_0 = 0$ corresponds to the constrained zero-temperature Glauber dynamics where the only possible moves are flips of isolated spins. The system, therefore, eventually reaches a blocked configuration, where there is no isolated spin \cite{god_luc_05}. 

In the previous paper we have decided to examine differences between both dynamics in one dimension under several types of updating \cite{szn_kru_06}. This issue seems to be quite important in the field of sociophysics, since several social experiments showed that updating plays an important role in opinion dynamics. For example, two groups of two people influence an individual stronger to conform than one group of four people \cite{Wilder77,Myers06}. 
On the other hand, updating schemes may play a prominent role in one-dimensional Ising model at zero temperature \cite{bolle_blanco_04,radicchi_etal_07}. 

Probably neither a completely synchronous nor a random sequential update is realistic for natural systems \cite{radicchi_etal_07}. Therefore, in \cite{szn_kru_06} we have introduced $c$-parallel updating (a randomly chosen fraction $c$ of spins is updated synchronously) and shown that outflow dynamics is much more influenced by the type of updating than inflow dynamics. The results in our previous paper were intriguing, but preliminary.  We looked at the mean relaxation time as a function of the initial fraction of randomly distributed up-spins only for several values of $c$ and $W_0$. Based on our previous simulations we noticed that:
\begin{enumerate} 
\item
Within $\frac{1}{L}$-parallel updating (i.e. random sequential updating) the relaxation is much slower for inflow  than for outflow dynamics. 
\item
Within $1$-parallel updating (i.e. deterministic synchronous updating) the relaxation under outflow is slower than under inflow dynamics. 
\item
In general, the relaxation times decay with $W_0$. 
\item
The dependence on $c$ is much stronger for outflow dynamics. For inflow dynamics the mean relaxation time is almost the same for all values of $c$. 
\end{enumerate}

As we show in this paper, the above results are valid only for some values of $W_0$ and $c$. Generally, the parameter space of our model is 3-dimensional and consists of: $c$ (fraction of spins updated synchronously), $W_0$ (probability of flipping a spin in lack of majority) and $p$ (initial fraction of up-spins). However, in this paper we focus on random initial conditions with equal number of up and down spins, i.e. $p=0.5$. This value of $p$ is the most frequently chosen in the computer simulations since it corresponds to high temperature situation in which magnetization is equal to 0. On the other hand, condition $p=0.5$ corresponds to the social situation in which public opinion is equal to zero, i.e. a democratic stalemate in which the society is not able to make any common decision. From such an initial condition, under one-dimensional zero-temperature outflow or inflow dynamics with random sequential updating, the system will eventually reach with equal probability one of two possible consensus states -- all spins up or all spins down. Obviously, the mean relaxation time for $p=0.5$ is larger than for any other value of $p$. For all these reasons the initial state with $p=0.5$ is the most interesting case. On the other hand, focusing on this particular value of $p$ allows us to limit the parameter space ($p,c,W_0$) to ($c,W_0$). In particular,  we show in this paper:
\begin{itemize}
\item
how the mean relaxation time $\tau$ scales with the system size $L$,
\item
what is the exact dependence between the mean relaxation time and parameters $c$ and $W_0$,
\item
what is the ratio between the mean relaxation time for inflow and for outflow dynamics as the function of parameters $c$ and $W_0$.
\end{itemize}

\section{Dynamics and updating}
Let us begin with recalling the definitions of inflow and outflow dynamics, as well as, the idea of the so called $c$-parallel updating. The system consists of $L$ Ising spins $S_i = \pm 1$ ($i=1,\ldots,L$) placed on the one-dimensional lattice with the periodic boundary conditions. 
We consider a quench from the infinite temperature to the zero temperature, and let
the system then evolve using one of two dynamics - the inflow and the outflow dynamics.

Generalized zero-temperature Glauber (inflow) dynamics can be defined without using the idea of the energy \cite{szn_kru_06}:
\begin{eqnarray}
\begin{array}{l}
S_i(t+1)  = \\
\left\{
\begin{array}{ll}
S_{i+1} (t) & \mbox{if }  S_{i-1}(t)S_{i+1}(t)=1, \\ 
-S_i(t) \hspace{0.2cm} \mbox{with prob} \hspace{0.2cm} W_0 & \mbox{if } S_{i-1}(t)S_{i+1}(t)=-1 \\ 
\end{array}
\right. 
\end{array}
\label{eq:in}
\end{eqnarray}

Similarly we can define the outflow dynamics. Recently \cite{sla_etal_08}, 
we have introduced slight modifications with respect to the original outflow rule:
choose pair of neighbors and if they both are in the same state, then 
adjust one (instead of two) of its neighbors (chosen randomly on left or right with
equal probability $1/2$) to the common state. Because this way at most one spin is flipped in one step, while in original formulation two can be flipped simultaneously, the time must be rescaled by factor $\frac{1}{2}$. We measure the time so that the speed of all processes remains constant when $L\to\infty$, so normally one update takes time $\frac{1}{L}$. Here, instead, we consider also the factor $\frac{1}{2}$, so single update takes time $\Delta t = \frac{1}{2L}$. Our modification eliminates some correlations due to simultaneous flip of spins at distance $3$. However, if we look at
later stages of the evolution , where typically the domains are larger
than $2$, simultaneous flips occurs very rarely. Therefore, we do not expect any substantial difference. Indeed, computer simulations confirmed our expectations - only time has to be rescaled. On the other hand, the modification simplifies the analytical treatment and allowed us to find exact formula for the exit probability \cite{sla_etal_08}.
Here we go even further and adjust left (instead of random) neighbor of chosen pair: 

\begin{eqnarray}
\begin{array}{l}
S_i(t+1)  = \\
\left\{
\begin{array}{ll}
S_{i+1} (t) & \mbox{if }  S_{i+1}(t)S_{i+2}(t)=1, \\ 
-S_i(t) \hspace{0.2cm} \mbox{with prob} \hspace{0.2cm} W_0 & \mbox{if } S_{i+1}(t)S_{i+2}(t)=-1 \\ 
\end{array}
\right. 
\end{array}
\label{eq:out}
\end{eqnarray}

\begin{figure}
\begin{center}
\includegraphics[scale=0.48]{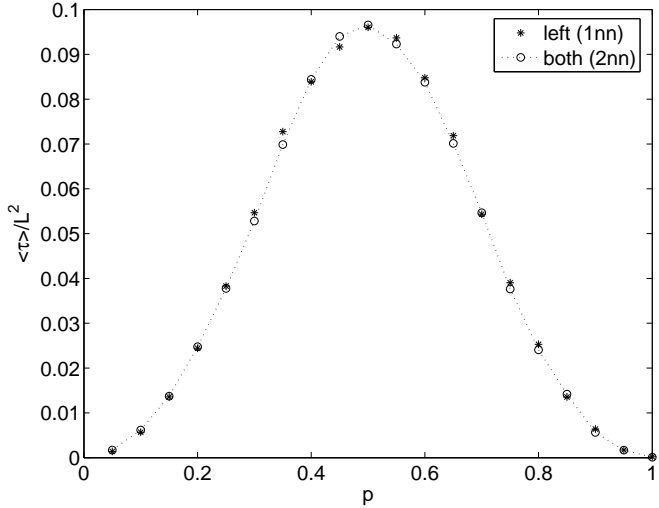}
\caption{The mean relaxation times from random initial state consisting of $p$ up-spins for the modified (1nn) and original (2nn) outflow dynamics in one dimension. In the modified version we adjust only left neighbor of chosen pair, so at most one spin is flipped in one elementary step while in original formulation two can be flipped simultaneously. Therefore in case of modified version the time was rescaled by factor $\frac{1}{2}$. It should be noticed that in computer simulations time is measured in Monte Carlo Steps (MCS). As usual, one MCS consist of $L$ elementary updating. Since we investigate the relaxation process we simulate the system as long as it reach the final state with all spins up or down. The average number of MCS needed to reach the final step depends on the initial concentration of up-spins and scales with the system size as $L^2$ analogously to the voter model \cite{L99,K92,MPR07}. The results presented on the plot are averaged over $10^4$ samples.}
\label{fig:fig0}
\end{center}
\end{figure}

We have checked by the computer Monte Carlo simulations that such a modification does not change results in the case of random sequential updating. Analogously to the case with one random neighbor \cite{sla_etal_08}, only the time has to be rescaled by factor $2$ in respect to the original rule in which two spins are changed in each elementary time step (see Fig. \ref{fig:fig0}). We introduce such a modification to simplify the case with synchronous updating. Moreover, it should be noticed that the case of synchronous updating for the zero-temperature Glauber dynamics is fully deterministic. By introducing modification described in Eq. (\ref{eq:out}) we avoid randomness in the case of synchronous updating for the outflow dynamics.

Seemingly there is only one small difference between both dynamics -- under inflow (see Eq. (\ref{eq:in})) spin $S_i$ is flipped according to its two nearest neighbors $S_{i-1}$ and $S_{i+1}$, and under outflow (see Eq. (\ref{eq:out})) according to its neighboring pair $S_{i+1},S_{i+2}$. This small difference has consequences even for random sequential updating - for $W_0=0$ we have constrained inflow dynamics where the only possible moves are flips of isolated spins and the system, therefore, eventually reaches a blocked configuration, where there is no isolated spin \cite{god_luc_05}. On the other hand, under outflow dynamics with $W_0=0$ the system will always eventually reach consensus or, in other words, ferromagnetic state. However, apart from the limit value of the flipping probability $W_0=0$ both dynamics seem to be qualitatively equivalent under random sequential updating. To see more differences between dynamics we have introduced  $c$-parallel updating \cite{szn_kru_06}. Within this updating a randomly chosen fraction $c$ of spins is updated synchronously. Of course, $c=1$ corresponds to deterministic synchronous updating and $c=1/L$ to random sequential updating.

\section{Simulation results}
\begin{figure}[htb]
\begin{center}
\includegraphics[scale=0.8]{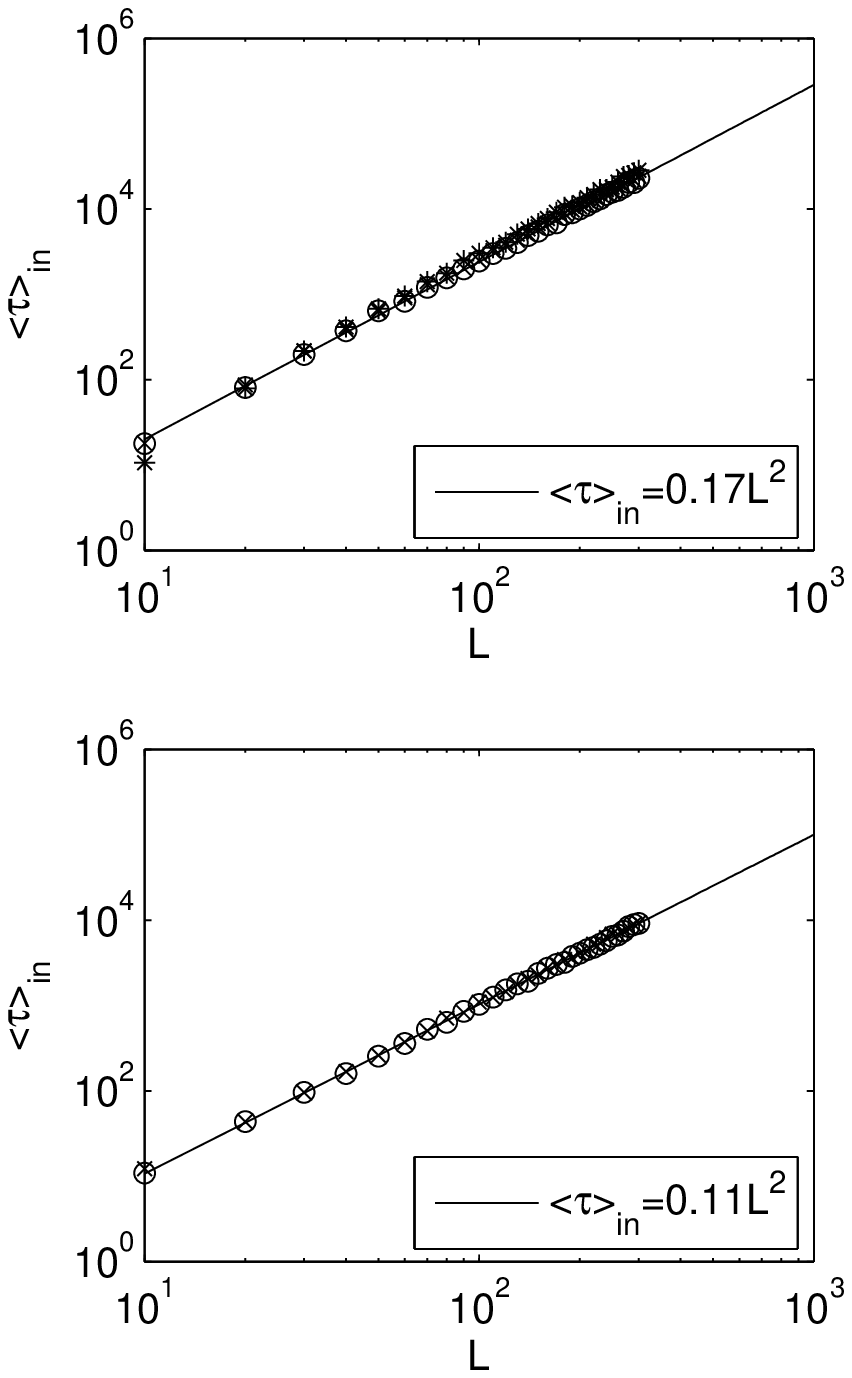}
\caption{The mean relaxation time from disordered to the final ferromagnetic steady state for one-dimensional \textbf{inflow dynamics} with $c$-paralel updating for $c=1$(*), $c=0.5$(x) and $c=1/L$(o) as a function of lattice size $L$. The mean relaxation time scales with the lattice size as $<\tau> \sim L^2$. The results presented on the plot are averaged over $10^3$ samples.}
\label{fig:scalin}
\end{center}
\end{figure}

In the previous paper \cite{szn_kru_06} we did not check how results scales with the system size $L$.
It is known that the mean relaxation time in one-dimensional system scales with the lattice size as $<\tau> \sim L^2$ for the voter model \cite{L99,K92,MPR07} as well as for the inflow and the outflow dynamics with random sequential updating \cite{szn_kru_06}. We decided to examine the scaling laws for $c$-synchronous updating with $c>1/L$. It occurs that in the case of inflow dynamics the mean relaxation time scales with the system size as $<\tau> = \alpha L^2$, where $\alpha=\alpha(W_0)$ and for a given value of $W_0$ is $c$-independent (see Fig. \ref{fig:scalin}). This results is valid for $W_0 < 0.5$. However, it has been shown recently that for $W_0=1$ there is a critical value of $c=\widetilde{c}$, above which the relaxation time is infinite -- the system never reaches the ferromagnetic steady state \cite{radicchi_etal_07}. Probably in general $\widetilde{c}=\widetilde{c}(W_0)$, but this is not the subject of this paper. For $W_0 < 0.5$ the system always eventually reaches the ferromagnetic steady state (i.e. $\widetilde{c}=0$ for $W_0 < 0.5$) and the scaling law $<\tau> = \alpha(W_0) L^2$ is valid.

\begin{figure}[htb]
\begin{center}
\includegraphics[scale=0.8]{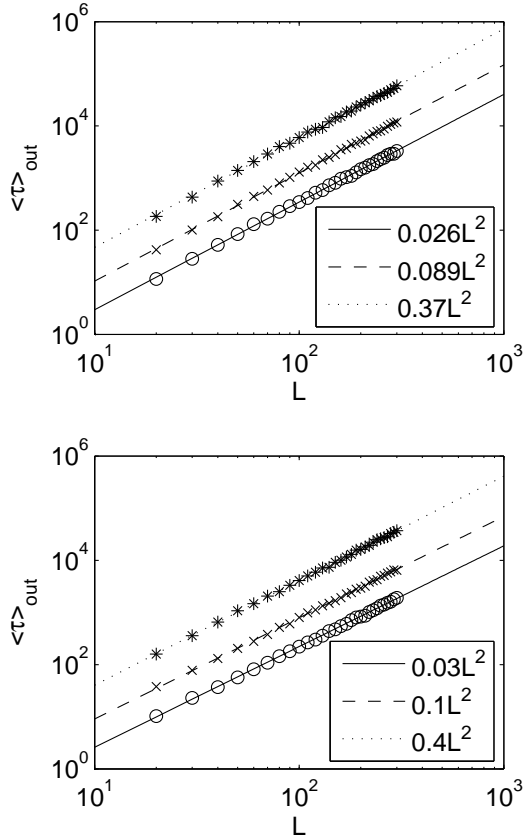}
\caption{The mean relaxation time from disordered to the final ferromagnetic steady state
for one-dimensional \textbf{outflow dynamics} with $c$-parallel updating for $c=1$(*), $c=0.5$(x) and $c=1/L$(o) as a function of lattice size $L$. The mean relaxation time scales with the lattice size as $<\tau> \sim L^2$. The results presented on the plot are averaged over $10^3$ samples.}
\label{fig:scalout}
\end{center}
\end{figure}

In the case of the outflow dynamics the mean relaxation time scales with the lattice size as $<\tau> \sim \alpha L^2$ and $\alpha=\alpha(c,W_0)$ (see Fig. \ref{fig:scalout}). Results presented in Figs. \ref{fig:scalin} and \ref{fig:scalout} confirm observation from our previous paper \cite{szn_kru_06} -- for $W_0 < 0.5$ outflow dynamics is more influenced by the type of updating than inflow dynamics. It should be noticed that $W_0$, with respect to social applications, can be interpreted as the probability of opinion change in case of uncertainty (lack of unanimity) and therefore it should be relatively small due to the observation that an individual who breaks the unanimity principle reduces social pressure of the group dramatically \cite{Myers06}. 

\begin{figure}[htb]
\begin{center}
\includegraphics[scale=0.45]{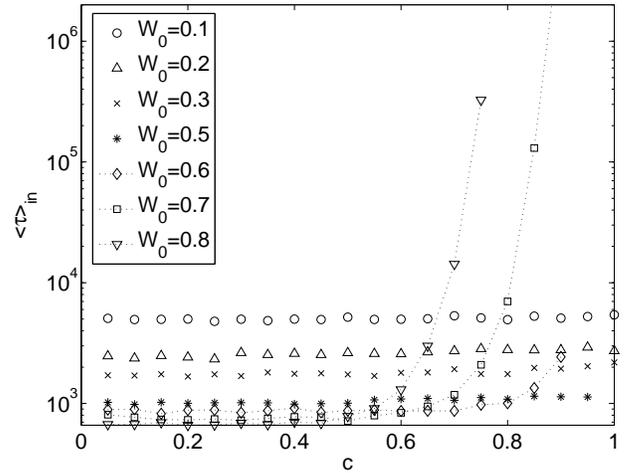}
\caption{The mean relaxation time from disordered to the final ferromagnetic steady state
for one-dimensional \textbf{inflow dynamics} with $c$-parallel updating as a function of $c$. For $W_0 < 0.5$ the mean relaxation time is $c$-independent. For $W_0>0.5$ the mean relaxation time increases rapidly with $c$ above certain value $c=c^*$. The results presented on the plot are averaged over $10^3$ samples for $L=100$.}
\label{fig:tauin}
\end{center}
\end{figure}

\begin{figure}[htb]
\begin{center}
\includegraphics[scale=0.45]{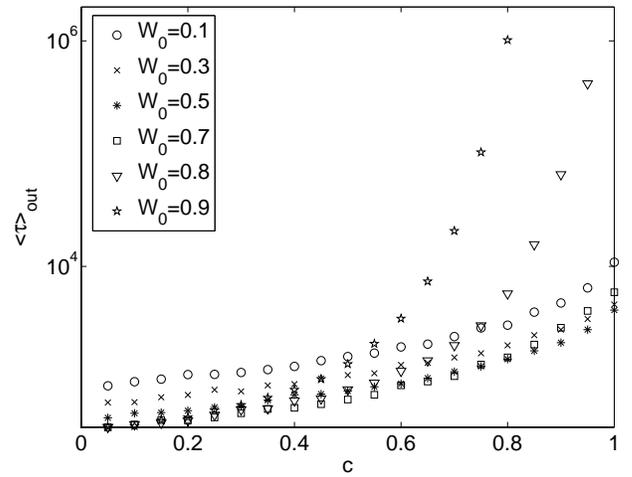}
\caption{The mean relaxation time from disordered to the final ferromagnetic steady state
for one-dimensional \textbf{outflow dynamics} with $c$-parallel updating as a function of $c$. In case this case the dependence on $c$ is visible for all values of $W_0$. The results presented on the plot are averaged over $10^3$ samples for $L=100$.}
\label{fig:tauout}
\end{center}
\end{figure}

Let us now examine the dependence between the mean relaxation time $<\tau>$ and the type of updating (characterized by $c$) for different values of flipping probabilities $W_0$. Several interesting phenomena are seen in Figs. \ref{fig:tauin} and \ref{fig:tauout}. In the case of the inflow dynamics (see Fig. \ref{fig:tauin}):
\begin{enumerate}
\item
The mean relaxation time $<\tau>$ is $c$-independent for flipping probability $W_0 < 0.5$. 
\item
For $W_0>0.5$ the mean relaxation time increases rapidly with $c$ above certain value $c=c^*$ (e.g. for $W_0=0.6, c^* \approx 0.7$, for $W_0=0.8, c^* \approx 0.45$). It has been found recently \cite{radicchi_etal_07} that for $W_0=1$ the system never reaches its ferromagnetic ground state for $c>\widetilde{c}=0.41$, (i.e. for $W_0=1$ and $c>0.41$ $<\tau> = \infty$). We expect that $c^*$ is correlated or even equivalent with critical $\widetilde{c}$. However, determining precise dependence between the critical updating scheme $c$ and flipping probability $W_0$ is not the subject of this paper and will be considered in a future study.
\item
Generally, the mean relaxation time decreases with increasing $W_0$. However, for $W_0>0.5$ this statement is valid only for $c<c^*$. 
\end{enumerate}  
In the case of the outflow dynamics (see Fig. \ref{fig:tauout}):
\begin{enumerate}
\item
The mean relaxation time $<\tau>$ depends on $c$ for all values of $W_0$. More precisely, the mean relaxation time increases with $c$, which is desired result in modeling social systems -- it was shown in series of social experiments that two groups of two people influence an individual stronger than one group of four people \cite{Wilder77,Myers06}.
\item
For $W_0 < 0.5$ the mean relaxation time decreases with increasing $W_0$ for whole range of $c$, analogously to inflow dynamics.
\end{enumerate}  

\begin{figure}[htb]
\begin{center}
\includegraphics[scale=0.45]{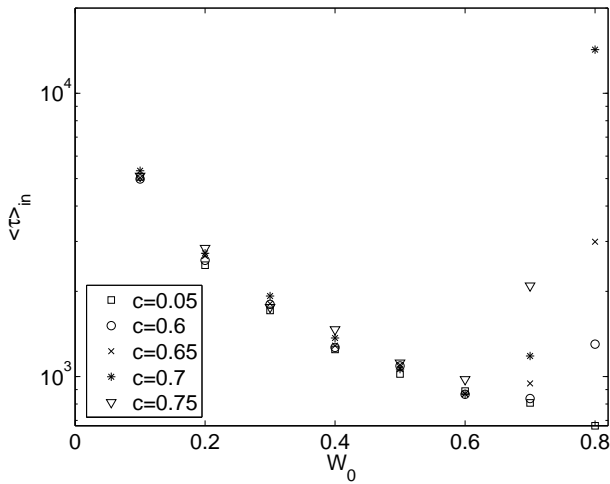}
\caption{The mean relaxation time from disordered to the final ferromagnetic steady state
for one-dimensional \textbf{inflow dynamics} with $c$-parallel updating as a function of the flipping probability $W_0$. The results presented on the plot are averaged over $10^3$ samples for $L=100$.}
\label{fig:tauwin}
\end{center}
\end{figure}

\begin{figure}[htb]
\begin{center}
\includegraphics[scale=0.45]{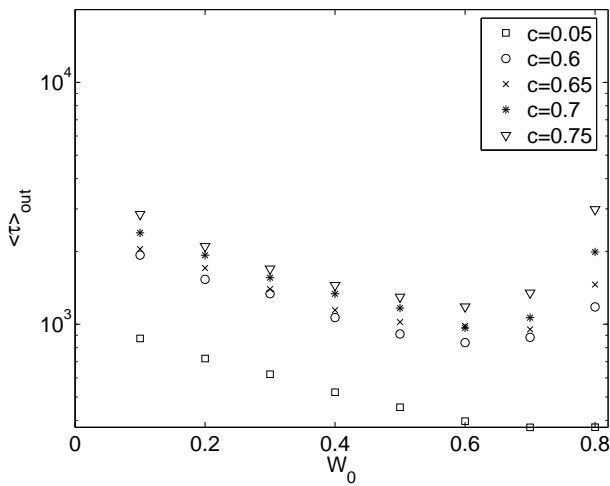}
\caption{The mean relaxation time from disordered to the final ferromagnetic steady state
for one-dimensional \textbf{outflow dynamics} with $c$-parallel updating as a function of the flipping probability $W_0$. The results presented on the plot are averaged over $10^3$ samples for $L=100$.}
\label{fig:tauwout}
\end{center}
\end{figure}

Let us now present dependence between the mean relaxation time and flipping probability $W_0$ (see Figs. \ref{fig:tauwin} and \ref{fig:tauwout}). In the previous paper it was argued that in general, the relaxation times decay with $W_0$ \cite{szn_kru_06}. However, this statement is valid only for $c \rightarrow \frac{1}{L}$. For larger values of $c$ dependence $<\tau(W_0)>$ is non monotonic for both (inflow and outflow) dynamics. The mean relaxation time decays with $W_0$ for $W_0<W_0^*(c)$ and grows above this value. This is very intriguing result but, due to partially synchronous updating scheme, we were not able to provide any analytical treatment that helps to understand this non-monotonic behavior. As we have already mentioned, we expect the critical value of $c^*(W_0)$, and analogously $W_0^*(c)$, above which the relaxation time is infinite. However, due to infinite relaxation (simulation) time, the critical values should be determined by measuring the evolution of active bonds instead of relaxation time. This will be studied in a future.

\begin{figure}[htb]
\begin{center}
\includegraphics[scale=0.45]{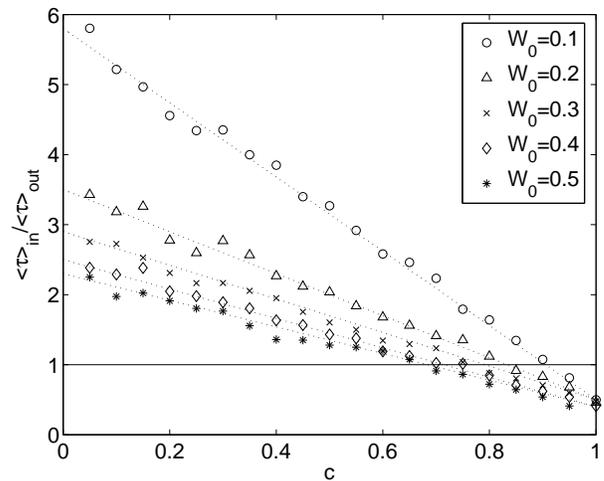}
\caption{The ratio between mean relaxation times for inflow and outflow dynamics as a function of $c$ for $W_0 \leq 0.5$. In this case the ratio $\tau_{in}/\tau_{out}$ decays linearly with $c$.}
\label{fig:ratio1}
\end{center}
\end{figure}

The main aim of this paper is to compare the mean relaxation times for the inflow and the outflow dynamics. Therefore we have decided to determine 
the ratio between mean relaxation times for inflow and outflow dynamics. For $W_0 \leq 0.5$ this ratio be can be approximated by linear function of updating scheme $c$ (see Fig. \ref{fig:ratio1}):
\begin{eqnarray}
\frac{<\tau>_{in}}{<\tau>_{out}}=-ac+\left( a+\frac{1}{2} \right),
\end{eqnarray}
where dependence between $a$ and $W_0$ can be approximated from simulation data (see Fig. \ref{fig:lin}) as:
\begin{eqnarray}
a=\alpha W_0 + 1 \hspace{0.5cm} \alpha \approx 0.43.
\end{eqnarray}
Finally:
\begin{eqnarray}
\frac{<\tau>_{in}}{<\tau>_{out}}=-\left( \alpha W_0 + 1 \right)c+\left( \alpha W_0 + 1.5 \right),
\end{eqnarray}
thus $<\tau>_{in}=<\tau>_{out}$ for:
\begin{equation}
c=\frac{\alpha W_0 + 0.5}{\alpha W_0 + 1} \hspace{0.5cm} (\alpha \approx 0.43)
\end{equation}
which in case of Glauber dynamics, i.e. $W_0=0.5$ gives $c \approx 0.48$.

\begin{figure}[htb]
\begin{center}
\includegraphics[scale=0.45]{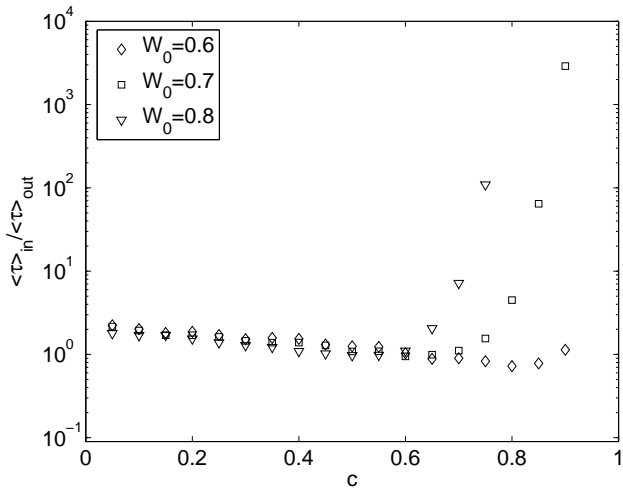}
\caption{The ratio between mean relaxation times for inflow and outflow dynamics as a function of $c$ for $W_0 > 0.5$. No linear dependence is seen in this case.}
\label{fig:ratio2}
\end{center}
\end{figure}

\begin{figure}[htb]
\begin{center}
\includegraphics[scale=0.45]{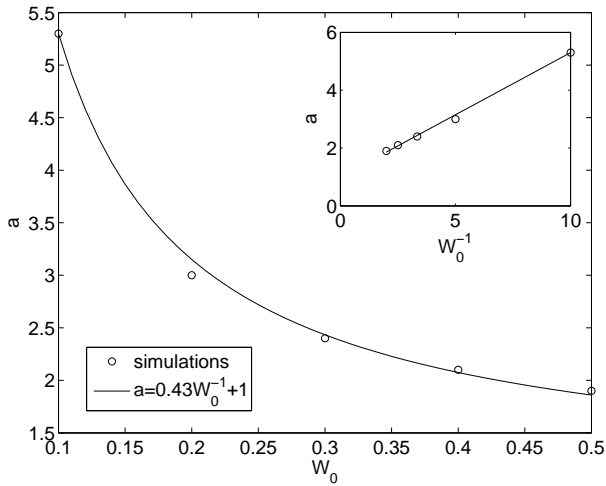}
\caption{For $W_0 \leq 0.5$ the ratio $\tau_{in}/\tau_{out}$ decays linearly with $c$  and it can be approximated as $\tau_{in}/\tau_{out}=-ac+(a+\frac{1}{2})$, where $a \sim W_0^{-1}$. Solid line represents power-law fit $a=0.43W_0^{-1}+1$. The same plot in log-log scale (inset) shows clearly the power-law dependence.}
\label{fig:lin}
\end{center}
\end{figure}

For $W_0>0.5$ ratio $<\tau>_{in}/<\tau>_{out}$ is non-monotonic (see Fig. \ref{fig:ratio2}). It decays with growing $c$ for $c<c^*(W_0)$ and above grows rapidly with $c$. Interestingly this special value of flipping probability $W_0=0.5$ corresponds to original Glauber dynamics \cite{Glauber_63}.

\section{Summary}
In this paper we examine the role of the so called $c$-parallel updating schemes in relaxation times from disordered states to the final ferromagnetic steady state.We investigate two zero-temperature single-spin flip dynamics on a one dimensional lattice of length $L$: inflow (i.e. generalized zero-temperature Glauber dynamics) and outflow opinion dynamics. Under random sequential updating, both dynamics seem to be equivalent in one dimension, except for one value of parameter $W_0 = 0$, which describes the probability of flipping the spin in lack of majority. 
Therefore, we have decided to examine differences between both dynamics in one dimension under generalized $c$-parallel updating. This issue seems to be particularly important in the field of sociophysics. In \cite{szn_kru_06} we have introduced $c$-parallel updating (a randomly chosen fraction $c$ of spins is updated synchronously). An analogous idea was proposed very recently by Radicchi et al. \cite{radicchi_etal_07} to study quench from infinite temperature to zero temperature of Ising spin systems evolving according to the Metropolis algorithm, in which the flipping probability associated with the $i$th spin is defined as \cite{mar_etal_94}:
\begin{equation}
p_M=\min \left\{ 1,\exp(-\Delta E_i/k_B T) \right\},
\end{equation} 
where $\Delta E_i$ is the energy change due to the $i$th spin flip. For $\Delta E_i=0$ we obtain $p_M=1$, hence zero-temperature limit of this algorithm can be rewritten as a special case of the so called generalized zero-temperature Glauber dynamics defined by Eq.(\ref{eq:in}), with $W_0=1$. 
It occurred that for $W_0=1$ the critical value of $c$ exists above which the system never reaches the final ferromagnetic state \cite{radicchi_etal_07}. This critical value was determined by measuring the evolution of active bonds. We expect that in the case of generalized Glauber dynamics ($W_0 \in [0,1]$) the critical value of $c$ depends on $W_0$. However, determining precise dependence between the critical updating scheme $c$ and flipping probability $W_0$ was not the subject of this paper and will be considered in a future study. In the present paper we focused on measuring the mean relaxation times $<\tau>$ to the ferromagnetic steady state as a function of $c$ and $W_0$ for inflow and outflow dynamics.

We have found the critical value of $W_0=0.5$ (which corresponds to the original Glauber dynamics \cite{Glauber_63}), below which the system eventually reaches the final ferromagnetic steady state with probability 1 for any value of $c$. 
Moreover, for $W_0 < 0.5$:
\begin{enumerate}
\item
the mean relaxation time scales with the system size as power law $<\tau> \sim L^2$ for both dynamics,
\item
the mean relaxation time decreases with increasing $W_0$ for both dynamics,
\item
the ratio between the mean relaxation times for inflow and outflow dynamics can be approximated by the linear function:
\begin{eqnarray}
\frac{<\tau>_{in}}{<\tau>_{out}}=-\left( \alpha W_0 + 1 \right)c+\left( \alpha W_0 + 1.5 \right),
\end{eqnarray}
\item
the mean relaxation time $<\tau>$ is $c$-independent for the inflow dynamics in contrast to the outflow dynamics.
\end{enumerate}  
For $W_0>0.5$, in the case of the inflow dynamics, the critical value of $c=c^*$ exists. Above this value the probability of reaching the final ferromagnetic state decreases to zero.  Simultaneously, the mean relaxation time increases rapidly with $c$ above the critical value $c=c^*$. 

Let us stress here, that for social applications results for $W_0 < 0.5$ are particularly important, since $W_0$ can be interpreted as the probability of opinion change in lack of majority. A number of social experiments showed that an individual who breaks the unanimity principle reduces social pressure of the group dramatically \cite{Myers06} and therefore we assume that $W_0$ is relatively small for social applications. Results obtained in this paper suggest that
the outflow dynamics is more adequate for modeling social phenomena than the inflow dynamics, since for $W_0 < 0.5$ the outflow dynamics is $c$-dependent in contrast to the inflow dynamics.  

On the other hand, inflow dynamics is much more adequate for physical applications. Moreover, it should be noticed that value $W_0=0.5$ corresponds to original Glauber dynamics \cite{{Glauber_63}}. Therefore, results obtained in this paper suggest that in some sense the original zero-temperature Glauber dynamics is a critical one among a broader class of inflow dynamics. The existence of the critical values of $c$ and $W_0$ in the $c$-parallel updating scheme for the generalized zero-temperature Glauber dynamics are very intriguing results and deserve deeper analysis.

\acknowledgments
We gratefully acknowledge the financial support of the Polish Ministery of Science and
Higher Education through the scientific grant no. N N202 0907 33


\begin{thebibliography}{33}
\bibitem{bolle_blanco_04}
D. Bolle and J. Busquets Blanco, Eur. Phys. J. B {\bf 42}, 397 (2004)

\bibitem{radicchi_etal_07}
F. Radicchi, D. Vilone, and H. Meyer-Ortmanns, Journal of Statistical Physics {\bf 129}, 593 (2007)

\bibitem{szn_kru_06}
K. Sznajd-Weron and S. Krupa, Phys. Rev. E {\bf 74}, 031109 (2006)

\bibitem{mar_etal_94}
A. M. Mariz, F. D. Nobre, C. Tsallis, Phys. Rev. B {\bf 49}, 3576 (1994)

\bibitem{god_luc_05}
C. Godreche and J. M. Luck, J. Phys.: Condens. Matter {\bf 17}, S2573 (2005)

\bibitem{Glauber_63}
R. J. Glauber, J. Math. Phys. {\bf 4}, 294 (1963)

\bibitem{szn_szn_00}
K. Sznajd-Weron and J. Sznajd, Int. J. Mod. Phys. C {\bf 11}, 1157 (2000)

\bibitem{KNC07}
D. T. Kenrick, S. L. Neuberg and R. B Cialdini, \emph{Social Psychology: Goals in Interaction (4th ed.)} Boston: Allyn and Bacon (2007)

\bibitem{AWA07}
E. Aronson, T. D. Wilson and R. M.  Akert  \emph{Social Psychology (6th ed.)} Englewood Cliffs, NJ: Prentice Hall (2007)

\bibitem{Myers06}
D. G. Myers, {\it Social Psychology}, McGraw-Hill Higher Education; 9 edition (2006)

\bibitem{Asch55}
S. E. Asch, Scientific American {\bf 193}, 31 (1955)

\bibitem{sta_etal_00}
D. Stauffer, A. O. Sousa, S. M. De Oliveira, Int. J. Mod. Phys. C {\bf 11}, 1239 (2000)

\bibitem{galam_86}
S. Galam, J. of Math. Psychology 30, 426 (1986) 

\bibitem{galam_90}
S. Galam, J. Stat. Phys. {\bf 61},  943  (1990)

\bibitem{kac_hol_96}
K. Kacperski, J. Ho{\l}yst, Journ. Statistical Physics, {\bf 84}, 169189 (1996)

\bibitem{hol_kac_sch_00}
J. Ho{\l}yst, K. Kacperski, F. Schweitzer, Physica A   {\bf 285}, 199 (2000)  

\bibitem{hol_kac_sch_01}
J.A. Ho{\l}yst, K. Kacperski and F. Schweitzer, Annual Review of Comput. Phys. {\bf 9}, 253 (2001)

\bibitem{kra_red_03}
P. L. Krapivsky and S. Redner, Phys. Rev. Lett. {\bf 90}, 238701 (2003)

\bibitem{gal_05}
S. Galam, Europhys. Lett., {\bf 70}, 705 (2005)

\bibitem{beh_sch_03}
L. Behera and F. Schweitzer,  Int. J. Mod. Phys. C {\bf 14} 1331 (2003) 

\bibitem{new_ste_99}
C. M. Newman, D. L. Stein, Phys. Rev. Lett. {\bf 82} 3944 (1999)

\bibitem{lipowski99}
A. Lipowski, Physica A {\bf 268}, 6 (1999) 

\bibitem{spi_kra_red_01b}
V. Spirin, P. L. Krapivsky, and S. Redner, Phys. Rev. E {\bf 65}, 016119 (2001)

\bibitem{spi_kra_red_01a}
V. Spirin, P. L. Krapivsky, and S. Redner, Phys. Rev. E {\bf 63}, 036118 (2001)

\bibitem{Jain02}
S. Jain, Physica A {\bf 305} 178 (2002) 

\bibitem{kra_red_tai_04}
P. L. Krapivsky, S. Redner and J. Tailleur, Phys. Rev. E {\bf 69}, 026125 (2004)

\bibitem{cha_das_08}
A. K. Chandra and S. Dasgupta, Phys. Rev. E {\bf 77}, 031111 (2008)

\bibitem{mor_etal_00}
A. A. Moreira, J. S. Andrade Jr., D. Stauffer, Int. J. Mod. Phys. C {\bf 12}, 39 (2001) 

\bibitem{ber_etal_01}
A. T. Bernardes, U. M. S. Costa, A. D. Araujo, D. Stauffer, Int. J. Mod. Phys. C {\bf 12}, 159 (2001) 
 
\bibitem{sta_oli_02}
D. Stauffer,P.M.C de Oliveira, Eur. Phys. J. B {\bf 30} 587 (2002) 

\bibitem{sta_02}
D. Stauffer, Comput. Phys. Commun. {\bf 146}, 93 (2002)

\bibitem{sch_02}
B. Schechter, New Scientist {\bf 175}, 42 (2002) 

\bibitem{for_sta_05}
S. Fortunato and D. Stauffer, in: Extreme Events
in Nature and Society, edited by S. Albeverio,
V. Jentsch and H. Kantz. Springer, Berlin - Heidelberg (2005)

\bibitem{sw_05}
K. Sznajd-Weron, Acta Physica Polonica B 36, 2537 (2005) 

\bibitem{cas_for_lor_07}
C. Castellano, S. Fortunato, V. Loreto, arXiv:0710.3256v1  

\bibitem{Wilder77}
D. A. Wilder, Journal of Experimental Social Psychology {\bf 13}, 253 (1977)

\bibitem{sla_etal_08}
F. Slanina, K. Sznajd-Weron, P. Przyby{\l}a, EPL {\bf 82}, 18006 (2008)

\bibitem{L99}
T. M. Liggett, \emph{Stochastic interacting systems: contact, voter, and exclusion
processes}, (Springer-Verlag, New York, 1999)

\bibitem{K92}
P. L. Krapivsky, Phys. Rev. A 45, 1067 (1992)

\bibitem{MPR07}
M. Mobilia, A. Petersen, S. Redner, J Stat Mech-theory E P08029 (2007)

\end{thebibliography}
\end{document}